\documentclass[aps,floats,twocolumn,showpacs]{revtex4-1}
\usepackage{amssymb}
\usepackage{amsmath}
\usepackage{epstopdf}
\usepackage{graphicx}

 \def\ds{\displaystyle}
 \def\bc{\begin{center}}          \def\ec{\end{center}}

 \allowdisplaybreaks
\begin{document}
 \title{Effect of plasma inhomogeneity on plasma wakefield acceleration driven by long bunches}
 \author{K.V.Lotov}
 \affiliation{Budker Institute of Nuclear Physics SB RAS, 630090, Novosibirsk, Russia}
 \affiliation{Novosibirsk State University, 630090, Novosibirsk, Russia}
 \author{A.Pukhov}
 \affiliation{Institut f\"ur Theoretische Physik I, Heinrich-Heine-Universit\"at D\"usseldorf,
40225 Germany}
 \author{A.Caldwell}
 \affiliation{Max-Planck-Institut f\"ur Physik, 80805, M\"unchen, Germany}
 \date{\today}
 \begin{abstract}
 Effects of plasma inhomogeneity on self-modulating proton bunches and accelerated electrons were studied numerically. The main effect is the change of the wakefield wavelength which results in phase shifts and loss of accelerated particles. This effect imposes severe constraints on density uniformity in plasma wakefield accelerators driven by long particle bunches. The transverse two stream instability that transforms the long bunch into a train of micro-bunches is less sensitive to density inhomogeneity than are the accelerated particles. The bunch freely passes through increased density regions and interacts with reduced density regions.
 \end{abstract}
 \maketitle

\section{Introduction}\label{s1}

Studies of plasma wakefield acceleration concentrate on various regimes of wave excitation depending on particle beam parameters. Early studies of plasma wakefield acceleration dealt with linear or weakly nonlinear wakes \cite{Chen,IEEE96} and were supported by proof-of-principle experiments with low density electron bunches \cite{JETPL13-354,FP20-663,PRL61-98,PRA39-1586,PFB2-1376,Jap,Jap2}. Then came the era of the blowout regime \cite{PRA44-6189}, stimulated by experiments at SLAC~\cite{PoP9-1845,Nat.445-741}. In parallel, interest in strongly nonlinear positron-driven wakes came up as intense positron bunches became available~\cite{PRL90-214801,PRL90-205002,PRL101-055001}.

Recently, the concept of proton-driven wakefield acceleration was introduced \cite{NatPhys9-363,PRST-AB13-041301} motivated by the huge particle energy and energy content of existing proton bunches. It was soon realized~\cite{PPCF53-014003} that it is much easier to excite the wave by a self-modulating long bunch rather than to compress the proton bunch longitudinally to sub-millimeter scales. If properly seeded, the transverse two-stream instability forms the bunch into equidistant micro-bunches spaced one plasma wavelength apart \cite{PRL104-255003,PoP18-024501}. The micro-bunches then resonantly drive the plasma wave to high amplitudes \cite{PoP18-103101}.

The process of resonant wave excitation assumes exact coincidence
of the plasma wavelength and the bunch-to-bunch distance. This requirement
is automatically met if the bunch self-modulates in a perfectly uniform
plasma. If the plasma density, $n$, is not constant along the line
of bunch propagation, then the plasma frequency $\omega_{p}=\sqrt{4\pi ne^{2}/m}$
and the wavelength $\lambda_{p}=2\pi c/\omega_{p}$ vary as well.
Here $e$ is the elementary charge, $m$ is the electron mass, and
$c$ is the light velocity. It has been shown in \cite{PRL107-145003}
that even a moderate density gradient of $0.5\%$ over one meter changes
the wake phase velocity drastically. This is due to the accumulation
of the wake phase shifts over the long proton bunch. The questions
`how precisely must the plasma uniformity be kept?' and `what are the
effects of possible non-unformities on the proton bunch modulation
and electron acceleration?' thus arise.

We address these questions by two-dimensional simulations of beam dynamics in plasmas with various density perturbations. In Section~\ref{s2} we define the problem, show the overall effect of plasma inhomogeneity on spectra of accelerated electrons, and study the effect of non-uniform plasmas on the proton bunch. In Section~\ref{s3} we look at electron dynamics and explain the observed results by shifts of the wakefield phase. In Section~\ref{s4} we summarize the main findings.

\section{Effect of density perturbations on the instability}\label{s2}

As a reference point for the study we take the half-cut proton bunch (SPS-LHC) discussed in \cite{PoP18-103101}. This case is one of possible scenarios for a proposed experiment on proton-driven wakefield acceleration at CERN. The initial bunch density is taken in the form
\begin{gather}\label{e1}
    n_b = \frac{n_{b0}}{2} e^{-r^2/2 \sigma_r^2} \left[  1 + \cos \left(  \sqrt{\frac{\pi}{2}} \frac{z}{\sigma_z}  \right)  \right], \\
    \label{e1a}
    n_{b0} = \frac{2 N_b}{(2\pi)^{3/2} \sigma_r^2 \sigma_z}, \quad  \sigma_z \sqrt{2\pi} < z < 0
\end{gather}
with the number of particles $N_b=5.75 \times 10^{10}$ and sizes $\sigma_r=0.2$\,mm and $\sigma_z=12$\,cm. Other important bunch parameters are the proton energy, $W_b= 450$\,GeV, and the emittance, 0.008\,mm\,mrad.

The nominal plasma density is $n_0 = 7 \times 10^{14}\text{cm}^{-3}$, so that $c/\omega_p \approx \sigma_r$. The bunch is thus very long if measured in plasma wavelengths: $\sigma_z \approx 100\,\lambda_p$.

The wakefield is sampled by a side injected low density electron bunch. The energy of electrons is 10\,MeV, and the sizes of the bunch are $\sigma_{e,r}=0.2$\,mm and $\sigma_{e,z}=5$\,mm.  The normalized emittance is 2.5\,mm\,mrad. The angle between the proton and electron bunches is 5\,mrad, and the beam axes cross at $z = 5.36$\,m (the proton bunch enters the plasma at $z=0$). This combination of injection position and angle was found to provide the best quality electron beam at the end of a 10\,m long plasma column. The electron bunch is timed to get trapped approximately 24\,cm, or $2 \sigma_z$, behind the leading edge of the proton bunch.

The evolution of the bunches in the plasma is simulated by the two-dimensional axisymmetric quasi-static code LCODE \cite{PoP5-785,PRST-AB6-061301} with the fluid solver for plasma electrons and the kinetic model for beam particles. The applicability of quasi-static codes  to studies of long bunches in short scale plasmas is discussed in Ref.\,\cite{PPCF52-065009}.

\begin{figure}[t]
 \bc\includegraphics[width=228bp]{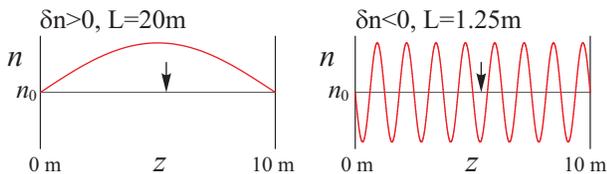} \ec
 \vspace*{-5mm}
\caption{(Color online) Examples of density perturbations tested. The arrows mark the location of the electron injection point.}\label{fig1-examples}
\end{figure}
\begin{figure}[t]
 \bc\includegraphics[width=231bp]{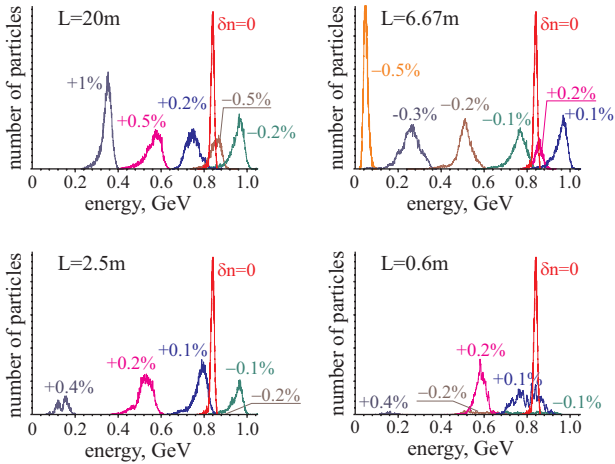} \ec
 \vspace*{-5mm}
\caption{(Color online) Spectra of accelerated electrons for density perturbations of various scale ($L$) and amplitude ($\delta n$).}\label{fig2-spectra}
\end{figure}
We first perturb the whole plasma column by taking its density in the form
\begin{equation}\label{e2}
    n = n_0 [1 + \delta n \sin (2 \pi z/L)]
\end{equation}
(Fig.\,\ref{fig1-examples}). The energy spectra  of accelerated electrons at $z=10$\,m are shown in Fig.\,\ref{fig2-spectra}. We see that rather small perturbations of the plasma density are sufficient to reduce the acceleration rate or even destroy the electron bunch. The shorter scale perturbations are more dangerous than the longer scale perturbations. The picture emerging from these spectra is rather complicated and results from a mixture of several effects.

\begin{figure*}[t]
 \bc\includegraphics[width=487bp]{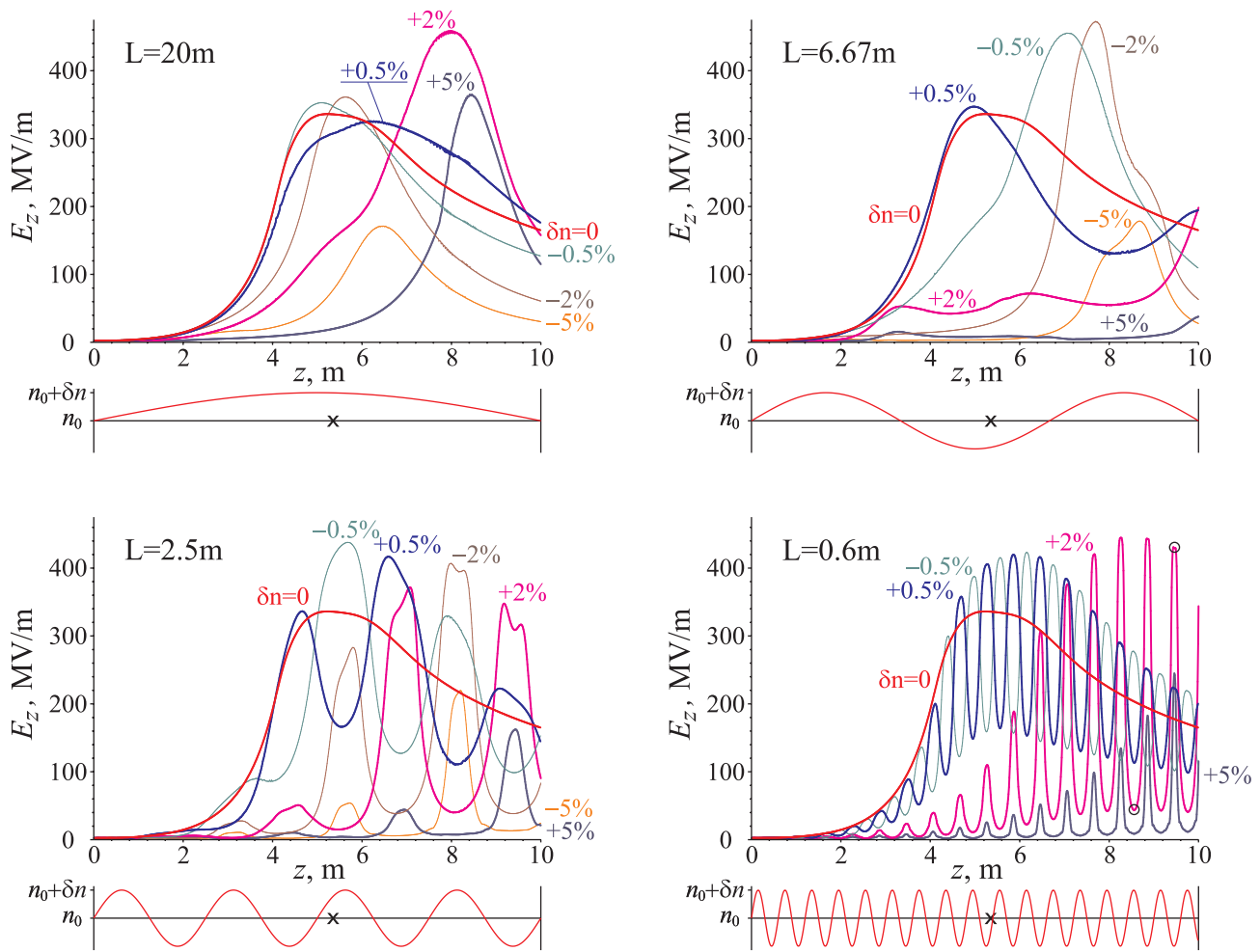} \ec
 \vspace*{-5mm}
\caption{(Color online) Maximum longitudinal electric field generated by the proton bunch in non-uniform plasmas versus the longitudinal coordinate~$z$. Insets under the graphs show corresponding profiles of the plasma density for positive $\delta n$. Crosses mark the electron injection point. The circles in the lower right plot indicate the cases shown in Fig.\,\ref{fig3a-field}.}\label{fig3-fields}
\end{figure*}
\begin{figure}[t]
 \bc\includegraphics[width=204bp]{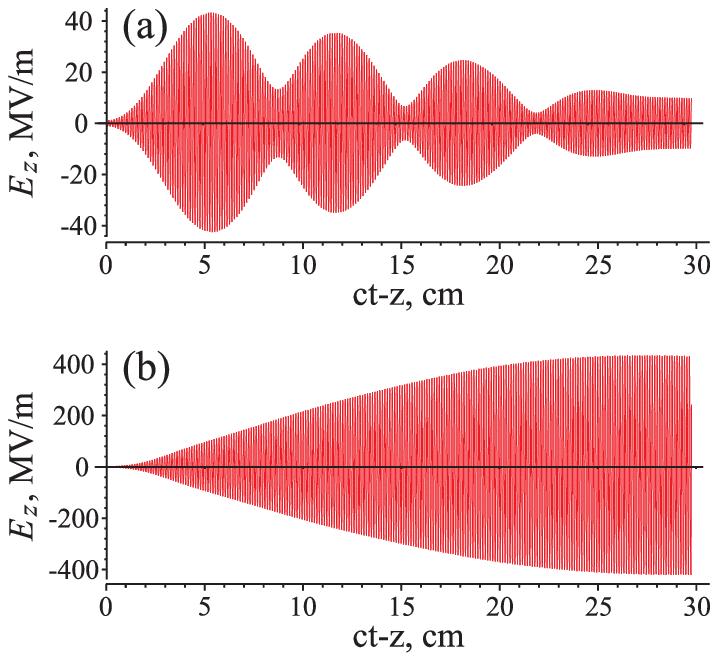} \ec
 \vspace*{-5mm}
\caption{(Color online) Longitudinal electric field in the non-uniform plasma with $L=0.6$\,m and $\delta n = 2$\% versus time at locations where (a) the wakefield is damped ($z=8.55$\,m) and (b) peaked ($z=9.45$\,m). Note the difference of vertical scales.}\label{fig3a-field}
\end{figure}
To characterize the effect of density perturbations on the proton bunch, we plot in Fig.\,\ref{fig3-fields} the maximum wakefield amplitude reached at different longitudinal locations in the plasma. We see that the instability is rather insensitive to regular density perturbations. Density variations as strong as 2\% do not suppress the instability, but just reduce the growth rate while keeping the saturated wave amplitude at approximately the same level.

Comparison of the wake amplitude patterns and density profiles show that the wake amplitude is largest in the regions of depressed plasma density, while in the regions with increased plasma density, the wake is damped.
The time dependence of the wakefield amplitude at fixed points in the plasma (Fig.\,\ref{fig3a-field}) indicates that the bunch is modulated at the plasma frequency of the lower density plasma regions.  When the modulated bunch enters the higher density regions, the modulation period is not resonant and a beating pattern emerges as seen in Fig.\,\ref{fig3a-field}a. This pattern is common to all simulated cases.

\begin{figure}[t]
 \bc\includegraphics[width=216bp]{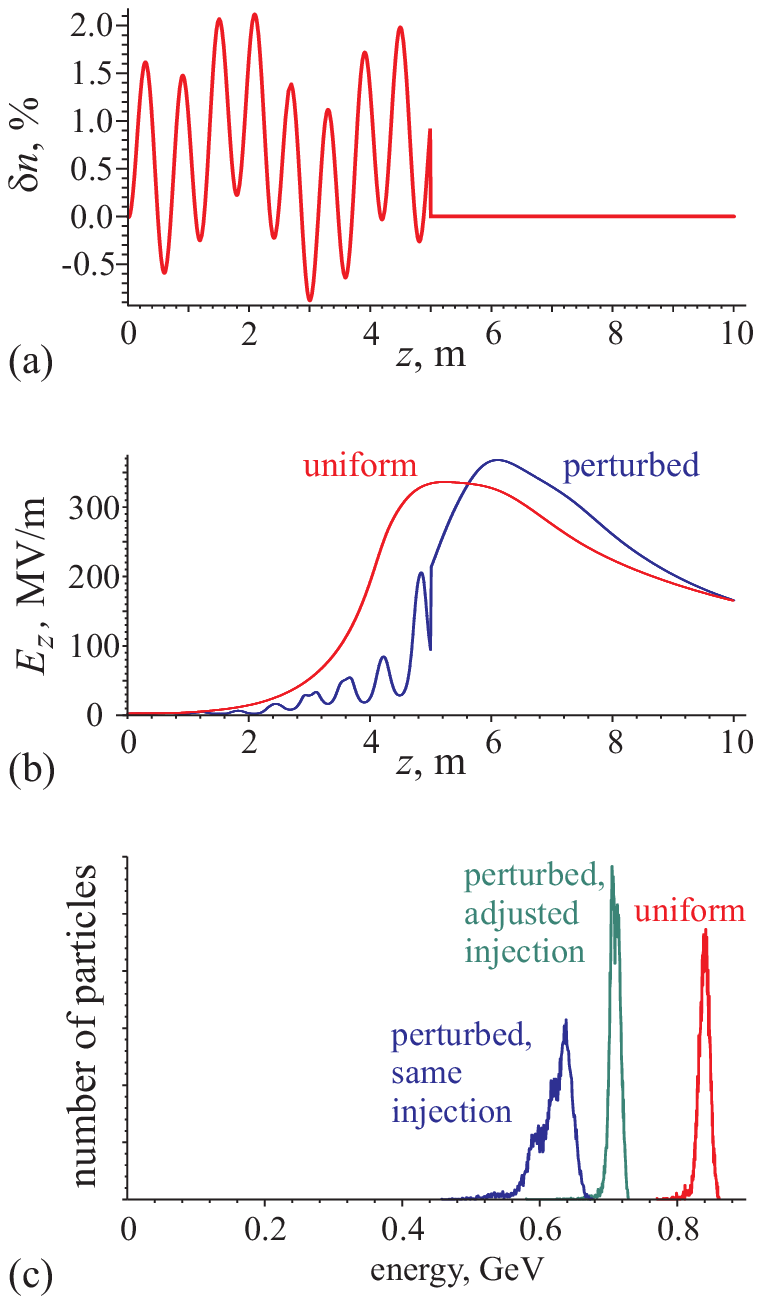} \ec
 \vspace*{-5mm}
\caption{(Color online) (a) Plasma density perturbation given by formula~(\protect\ref{e6}), (b) longitudinal electric fields generated in the uniform and this non-uniform plasmas, and (c) corresponding spectra of accelerated electrons at $z=10$\,m.}\label{fig4-start}
\end{figure}
It is instructive to compare the simulation results with the theory of instability growth in inhomogeneous plasmas \cite{PoP19-010703}. Two regimes of the instability are distinguished. For the long scale perturbations with $2\pi z/L \equiv k_f z \ll 1$, the instability is predicted to be suppressed for
\begin{equation}\label{e4}
    \delta n \, k_f \gtrsim \left(  \frac{n_{b0}m c^2}{2 n_0 W_b z \xi^2}  \right)^{1/3},
\end{equation}
where $\xi = z-ct$ is the co-moving coordinate measured from the bunch head. Our longest perturbations with $L=20$\, marginally fit this regime for $z \lesssim 3$\,m. For this $z$ and $\xi=-24$\,cm, formula (\ref{e4}) gives the cutoff value $\delta n = 0.6$\%. If we take into account that the on-axis bunch density is not constant between $\xi=0$ and $\xi=-24$\,cm and gradually decreases from $n_{b0}$ to some value, then the cutoff perturbation will be even lower. However, in simulations we observe almost unimpeded growth of the wakefield for $\delta n = \pm 2$\%. In the opposite case of short scale inhomogeneity, the cutoff perturbation is
\begin{equation}\label{e5}
    \delta n \sim \left(  \frac{n_{b0}m c^2 z^2}{2 n_0 W_b \xi^2}  \right)^{1/3}.
\end{equation}
Substituting $z=5$\,m and $\xi=-24$\,cm into this expression yields $\delta n = 0.8$\%, which is again well below values ($\pm 5$\%) for which the instability is suppressed in simulations.

\begin{figure*}[t]
 \bc\includegraphics[width=458bp]{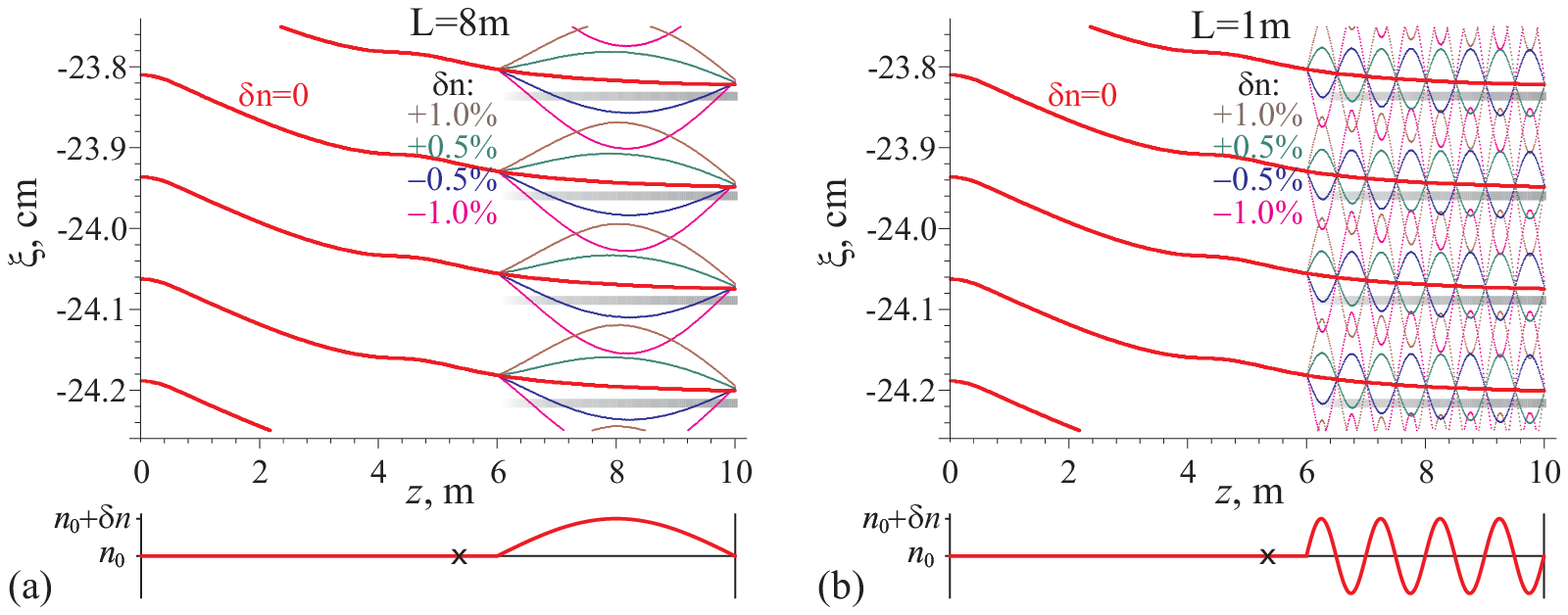} \ec
 \vspace*{-5mm}
\caption{(Color online) Locations of constant phase points (where $E_z=0$ on axis and $\partial E_z/\partial \xi >0$) versus propagation distance for the uniform (thick line) and perturbed (thin lines) plasmas. Grey bars show locations of trapped electrons. Insets under the graphs show corresponding profiles of the plasma density for positive $\delta n$.}\label{fig5-phase}
\end{figure*}
\begin{figure*}[t]
 \bc\includegraphics[width=493bp]{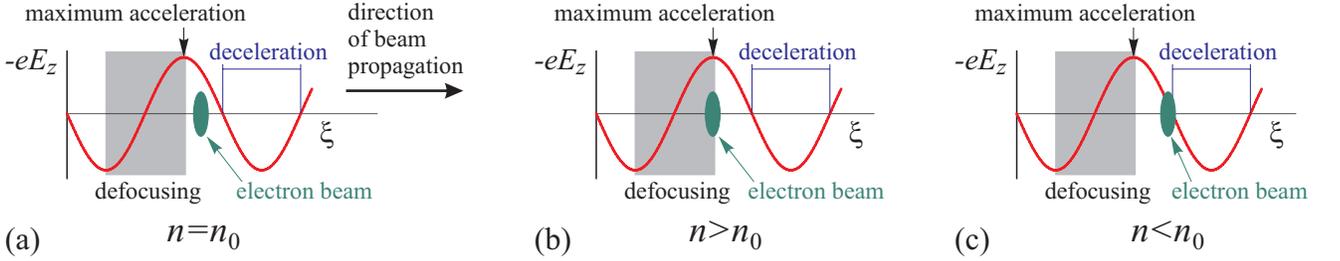} \ec
 \vspace*{-5mm}
\caption{(Color online) Phasing of the accelerated electron bunch in plasmas of the proper density (a), increased density (b), and reduced density (c). }\label{fig6-bunch}
\end{figure*}
The micro-bunches formed in an inhomogeneous plasma can be well suited for further wakefield excitation and electron acceleration. To illustrate this, we take the plasma density of the form
\begin{equation}\label{e6}
   n = \left\{ \begin{array}{llcl}
    n_0 \biggl[\!\!\!\!\!\!&\ds 1+0.01 \left(1 - \cos\left( \frac{2 \pi z}{\text{0.6\,m}}\right)\right) &&\\
        &\ds - 0.005 \sin \left( \frac{2 \pi z}{\text{2.5\,m}}\right)  &&\\
        &\ds - 0.005 \sin \left( \frac{2 \pi z}{\text{20\,m}}\right)\biggr], &  & z < \text{5\,m}, \\
    n_0, &&& z > \text{5\,m}.
                \end{array}
     \right.
\end{equation}
This density perturbation (Fig.\ref{fig4-start}a) contains spacial harmonics of different scales in the first half of the plasma cell to model the plasma inhomogeneity. In the second half, the density is kept uniform to distinguish the effects from plasma inhomogeneities on the proton and electron bunches. The instability still develops (Fig.\ref{fig4-start}b), and spectra of accelerated electrons are almost as good as for the uniform case (Fig.\ref{fig4-start}c). This numerical experiment also showed the importance of a proper injection point and angle. Adjustment of these two parameters was necessary to provide the best result for the particular instability case.

\section{Effect of density perturbations on accelerated electrons}\label{s3}

Plasma nonuniformity can affect the quality of the accelerated electron bunch in three ways: by changing proton bunch dynamics and the instability growth, by changing trapping conditions, and by changing accelerating properties of the wakefield. The instability growth has been analyzed in the previous section and was shown to be relatively insensitive to density variations. Trapping of side injected electrons, even in the uniform plasma, is a complicated process deserving a separate study, so we assume the electron bunch is already trapped and examine the acceleration only. We note that the trapping process produces a series of microbunches of electrons spaced at the plasma wavelength where the trapping has occurred.  To understand the effect of changing plasma density on these microbunches, we perturb the downstream part of the plasma column:
\begin{equation}\label{e7}
    n = \left\{ \begin{array}{lcl}
                  n_0, &  & z < z_0, \\
                  n_0 [1 + \delta n \sin (2 \pi (z-z_0)/L)], & \quad & z > z_0
                \end{array}
     \right.
\end{equation}
with $z_0 = 6$\,m. In the first 6 meters of the uniform plasma, the self-modulation fully develops and electrons are trapped and pre-accelerated to the energy of about 140\,MeV.

The main effect of density nonuniformity comes from dephasing of the wakefield and electron bunches. To illustrate this, we show in Fig.\,\ref{fig5-phase} how the wave phase moves with respect to the speed-of-light frame or, equivalently, to highly relativistic electrons. Lines in Fig.\,\ref{fig5-phase} are coordinates of the ``stable points'' where $E_z=0$ on axis and the wakefield is focusing for electrons. Only the wakefield fragment near the electron bunches is shown. In the first 4--5 meters of the plasma, the phase moves backward as expected for the developing instability \cite{PRL107-145003,PRL107-145002} and the wave is unusable for electron acceleration. After the bunch fully self-modulates at $z \sim 5$\,m, the phase velocity in the uniform plasma approaches light velocity, and the wave can accelerate electrons. In nonuniform plasmas, the phase responds to density variations by forward or backward shifts. This can be fatal for the accelerated bunches, as illustrated by Fig.\,\ref{fig6-bunch}. In the unperturbed plasma, electrons are located somewhere in the region of both focusing and acceleration [Fig.\,\ref{fig6-bunch}(a)]. If the plasma density increases, the wavelength shortens, the wave near the electron bunch shifts forward, and the bunch either enters the stronger accelerating field or falls into the defocusing region and gets scattered [Fig.\,\ref{fig6-bunch}(b)]. If the plasma density decreases, the wavelength elongates, and the bunch sees a lower accelerating field or even a decelerating field [Fig.\,\ref{fig6-bunch}(c)].

\begin{figure}[htb]
 \bc\includegraphics[width=229bp]{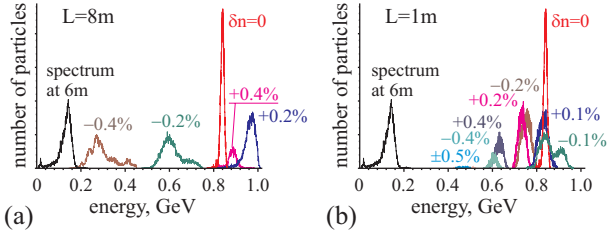} \ec
 \vspace*{-5mm}
\caption{(Color online) Spectra of accelerated electrons for the perturbed downstream part of the plasma column.}\label{fig7-spectra}
\end{figure}
Energy spectra of electrons at $z=10$\,m are consistent with this model. For long scale non-uniformities [$L=8$\,m, Fig.\,\ref{fig7-spectra}(a)], the density perturbation has a definite sign, and we observe either monotonous reduction of the beam energy as $|\delta n|$ increases (for $\delta n<0$), or some growth of the beam energy followed by complete loss of electrons (for $\delta n>0$). For short scale non-uniformities [$L=1$\,m, Fig.\,\ref{fig7-spectra}(b)], effects of different signs of $\delta n$ are averaged, and we observe a simultaneous decrease of the electron energy and population.

\begin{figure}[htb]
 \bc\includegraphics[width=203bp]{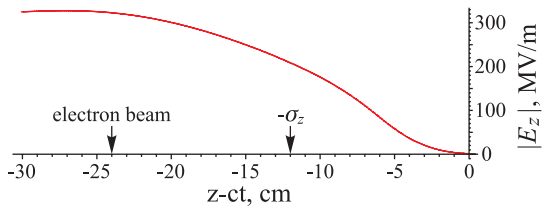} \ec
 \vspace*{-5mm}
\caption{(Color online) Instant snapshot of the wakefield amplitude for $ct=5$\,m.}\label{fig8-field}
\end{figure}
To find out the acceptable magnitude of density perturbations $\delta n_\text{max}$, we refer to the snapshot of the wakefield amplitude at the stage of the fully modulated proton bunch (Fig.\,\ref{fig8-field}). There are two important points on this graph. One is the location of the electron bunch ($\xi \approx -24$\,cm). Another is the place in which the wakefield is mainly formed, or more strictly the point where the field has a strength of half of the field that accelerates electrons ($\xi \sim -\sigma_z =  -12$\,cm). The distance between the two is the ``length'' of the useful wakefield. Let $N_\text{per}$ be this distance measured in plasma wavelengths. In our case $N_\text{per} \approx 100$. Since $\lambda_p \propto n^{-1/2}$, then the relative change $\delta n$ of the plasma density results in the relative change $\delta \lambda_p = -\delta n/2$ of the plasma wavelength and shift of the wakefield phase by the fraction $N_\text{per} \delta n/2$ of the period. If the electron bunch is located in the middle of the focusing and acceleration interval, as is shown in Fig.\,\ref{fig6-bunch}(a), then the shift of the wave by 1/8 of the period will bring the bunch to a defocusing or decelerating phase. We consider this shift as the maximum acceptable one. Thus we come to the engineering formula
\begin{equation}\label{e8}
   \delta n_\text{max} \sim 0.25/N_\text{per}.
\end{equation}
Here $\delta n_\text{max} \sim 0.25$\% in agreement with Fig.\,\ref{fig7-spectra}.

\begin{figure}[htb]
 \bc\includegraphics[width=219bp]{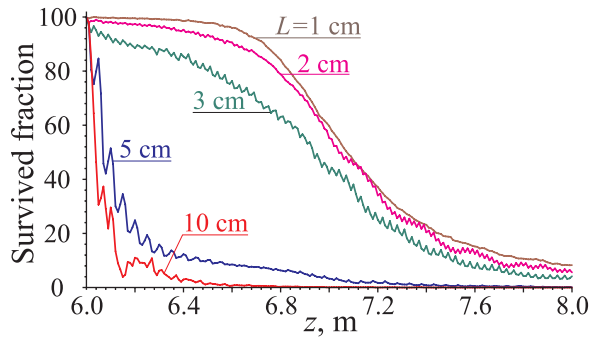} \ec
 \vspace*{-5mm}
\caption{(Color online) Fraction of the electron bunch remaining in the circle of radius $\sigma_r$ for strong plasma nonuniformities of different length scales $L$.}\label{fig9-loss}
\end{figure}
If the density perturbation is much stronger than the acceptable value (\ref{e8}), then the electron bunch is scattered away at the defocusing distance
\begin{equation}\label{e9}
    L_d = \frac{c}{\omega_p} \sqrt{\frac{\gamma_e}{\tilde E}},
\end{equation}
where $\gamma_e$ is the relativistic factor of the electrons, and $\tilde E$ is the wave amplitude in units of the wavebreaking field $m c \omega_p/e$. The distance (\ref{e9}) is also the maximum scale of the increased plasma density region which the electron bunch can pass through. In the case considered, we have $\tilde E \approx 0.1$ at $z=6$\,m. Thus 10\,MeV electrons can survive in plasma nonuniformities up to 0.25\,cm long. For 140\,MeV electrons this length is 1\,cm, and for 1\,GeV electrons 2.5\,cm. The distances are quite short, so keeping the plasma uniformity better than (\ref{e8}) is a must once electrons are injected into the wakefield. Figure~\ref{fig9-loss} illustrates how quickly electrons are lost for plasma perturbations (\ref{e7}) with $\delta n = 0.01 > \delta n_\text{max}$ and various $L$. For $L>L_d$, most of the bunch is scattered by the first density maxima. For $L \lesssim L_d$, the bunch survives over much longer distance under the action of some averaged focusing force.

\section{Summary}\label{s4}

Plasma inhomogeneities result in varying frequencies of the plasma oscillations, causing phase differences between the driving force (the modulated proton bunch) and the plasma wave, as well as phase differences between the accelerated particles and the wakefields. If accelerated particles are located $N_\text{per}$ wakefield periods behind the location of wakefield formation, then relative density perturbations stronger than (\ref{e8}) are sufficient to destroy the accelerated bunch at the defocusing distance (\ref{e9}). This effect imposes severe constraints on density uniformity in plasma wakefield accelerators driven by long particle bunches.

The transverse two stream instability that transforms the long drive bunch into a train of micro-bunches is less sensitive to density inhomogeneity than are accelerated particles. The drive bunch freely passes through increased density regions and interacts with reduced density regions. However, the spacing of the micro-bunches produced in this case corresponds to a plasma density which is lower than the path average.

\section{Acknowledgements}

This work is supported by the Ministry of Education and Science of the Russian Federation, RFBR (grants 11-01-00249, 11-02-00563, and 12-01-00727), grant 11.G34.31.0033 of the Russian Federation Government, and RF President's grant NSh-5118.2012.2.

\end{document}